\documentclass[preprint,12pt]{elsarticle}
\usepackage{graphicx}
\usepackage{amssymb}
\usepackage{amsthm}
\usepackage{bm}   
\usepackage{color}
\usepackage{amsmath}
\usepackage{balance}
\usepackage{subfigure}
\usepackage{relsize}
\usepackage{float}
\newcommand{\mathsym}[1]{{}}
\newcommand{\unicode}[1]{{}}
\usepackage{setspace}

\begin{document}
	\title{\textbf{ Degenerate soliton solutions and their interactions in coupled Hirota equation with trivial and nontrivial background}}
	
	\author[inst1]{S. Monisha} 
 \address[inst1]{ Department of Nonlinear Dynamics, Bharathidasan University,\\Tiruchirappalli - 620024, Tamil Nadu, India.}

\author[inst2]{N. Vishnu Priya}
\address[inst2]{Department of Mathematics, Indian Institute of Science,\\ Bangalore - 560012, Karnataka, India.} 
\author[inst1]{M. Senthilvelan}

	\begin{abstract}We construct two kinds of degenerate soliton solutions, one on the zero background and another on the plane wave background for the coupled Hirota equation. In the case of zero background field, we derive  positon solutions of various orders.  We also study interaction dynamics between positon solutions through asymptotic analysis and show that the positons exhibit time dependent phase shift during collision.  We also construct hybrid solutions which composed of positons and solitons and examine the interaction between higher order positon and multi-solitons in detail.  From the interaction, we demonstrate that the occurrence of elastic and inelastic interaction between multi-solitons and higher order positons. Further, we construct bound states among solitons and positons for the coupled Hirota equation.  In the case of plane wave background, we construct breather-positon solutions. For the coupled Hirota equation, the breather-positon solutions are being reported first time in the literature.  From the breather-positon solutions, we bring out certain interesting collision dynamics between breather-positons and positons. \end{abstract} 
\maketitle
\section{Introduction}
 The properties of nonlinear localized waves such as solitons, breathers and rogue waves play a  predominant role in the field of plasma physics, fluid mechanics, water waves, condensed matter physics and nonlinear optics \cite{1,Akhm,3,4}.  At first, a lot of attention has been paid to obtain soliton solutions which describe the solitary wave phenomena in shallow water surfaces. Subsequently, interest has been shown in exploring other kinds of localized solutions including breathers and rogue waves in integrable nonlinear evolutionary equations. In recent years, exploring degenerate solutions in nonlinear integrable systems gained momentum among the research community. In this direction, very recently one such degenerate soliton solution, namely positons is being studied widely in the nonlinear dynamics literature. \\
 \par The method of finding positon solution for the nonlinear partial differential equation (PDE) was first introduced by Matveev for the Korteweg-de Vries (KdV) equation \cite{M1,m2} where it results as a singular positon solution. This type of solution may model the shallow water rogue waves \cite{RW}.  Thereafter the procedure was extended to  hierarchies of KdV equations \cite{kdv1,kdv2,kdv3} and sine-Gordon equation \cite{SG}. Subsequently, attempts have been made to derive  non-singular positon solution for the nonlinear evolutionary equations and the resultant solutions are termed as smooth positons in the literature.  Smooth positon solutions may represent the Tidal-bore phenomenon in which the waves travel more than hundred kilometers with same amplitude and velocity \cite{tidal}.  These smooth positon solutions have been constructed for several nonlinear evolutionary equations including nonlinear Schr\"{o}dinger (NLS) equation \cite{sonls}, derivative NLS equation \cite{dnls}, complex mKdV equation \cite{cmkdv}, Chen–Lee–Liu equation \cite{CLL}, to name a few.  Positon solutions are also constructed on the plane wave background which in turn gave birth to breather positon (b-p) solution.  These b-p solutions are also derived and analyzed for few equations such as Sasa–Satsuma equation\cite{bp-ss}, complex mKdV equation \cite{cmkdv},  NLS–Maxwell Bloch equations \cite{bp-nls-mb}, Kundu–Eckhaus equation \cite{bp-KE}.  Recently, we have derived positon and b-p solutions for a generalized NLS equation \cite{w1} and the extended NLS equation \cite{w2}.  All these equations describe only a single amplitude wave which subject to different nonlinear effects.  
    \par Differing from the above, in the description of bire-fringent optical fibers, financial field and Bose-Einstein condensates one comes across more than single amplitude wave which in turn constitute coupled nonlinear PDEs.  These coupled equations explain the real world problems more accurately than a single amplitude wave equation.  In the literature, the study on positons is not well established for coupled equations.  Recently, positon solutions have been studied for the Manakov system and the interaction between solitons and positons are  also discussed briefly in \cite{cnls}. Breather positon solutions of coupled NLS equation is yet to be reported.  Motivated by this, in this work, we consider a coupled NLS equation, which has higher order dispersion and nonlinear effects, namely coupled Hirota (CH) equation and derive higher order positon solutions, b-p solutions and analyze their interactions with bright and dark solitons. We consider the CH equation in the form
    \begin{subequations}
		\begin{equation}
			i \psi_t +\dfrac{1}{2} \psi_{xx} + \left(|\psi|^2 + |\phi|^2\right)\psi + i \alpha \left(\psi_{xxx} +\left(6 |\psi|^2 + 3|\phi|^2\right)\psi_x + 3 \psi \phi^* \phi_x\right)=0,
		\end{equation}
		\begin{equation}
			i \phi_t +\dfrac{1}{2} \phi_{xx} + \left(|\psi|^2 + |\phi|^2\right)\phi + i \alpha \left(\phi_{xxx} +\left(6 |\phi|^2 + 3|\psi|^2\right)\phi_x + 3 \phi \psi^* \psi_x\right)=0, 
		\end{equation}
	\end{subequations}\label{e1}
    where $\psi(x,t)$ and $\phi(x,t)$ are the complex wave envelope functions and $\alpha$ is a real parameter. The higher order nonlinear effects are essential in the field of birefringent or two-mode nonlinear fibers in order
    to describe the propagation of femtosecond optical pulses. When $\alpha =0$, Eq.\eqref{e1} reduces to the celebrated Manakov system \cite{manakov}. Tasgal and Potasek included certain high-order effects like third order dispersion, self-steepening and inelastic Raman scattering terms in the CH equation \cite{TP}. Equation \eqref{e1} is an integrable equation which can be confirmed from the presence of Lax pair \eqref{lp} which was reported in \cite{TP}. Further, Painlev\'{e}  analysis and Darboux transformation method have also been applied for the Eq.\eqref{e1} in Refs. \cite{ds-ch,os}. Studies on localized waves solutions such as bright and dark soliton, dark rogue wave and composite rogue wave solutions for the Eq.\eqref{e1} can be found in Refs. \cite{rw-ch,rw-ch2}. 
  \par The positon solutions of CH Eq. \eqref{1} is yet to be reported in the literature. In this work, we construct two kinds of degenerate solutions for the Eq. \eqref{e1} using generalized Darboux transformation (GDT) method \cite{Matveevbook,DT-che} - one on the zero background and another on the constant (plane wave) background.  In the case of zero background field we derive two types of solutions, namely (i) smooth positon solutions of various orders and (ii) hybrid solutions composed of solitons and smooth positons.  In the case of smooth positons we derive second, third and fourth order smooth positon solutions for the Eq.\eqref{1}.  We also analyze the asymptotic behaviour of second order positon solutions of \eqref{e1}. In the asymptotic limits, positon exhibits time dependent phase shift.  We derive an exact expression for the phase shift and demonstrate that the displacement function/phase shift depends on the system parameter ($\alpha$) which influences the smooth positons by enlarging the distance between them. This is in contrast with solitons which exhibit constant phase shift in their asymptotic limits \cite{kdv, SineG,Hirota2}.  We then move on to construct hybrid solutions and study the interaction dynamics between higher order positon and soliton solutions of \eqref{e1}. In particular, we examine the interaction between (i) second order positon and one soliton, (ii) second order positon and two soliton and (iii) third order positon and one soliton. For certain parameter values, we come across elastic interaction between soliton and positon solution \cite{che}. In addition to the above, we construct bound states among soliton and positon for the Eq.\eqref{e1}.
 
 \par  With nonzero constant background, we derive degenerate breathers/ breather positon solutions.  To the best of our knowledge, breather positon solution  for a coupled nonlinear PDE is being reported first time in the literature which involves lengthy and tedious calculations.  Upon choosing plane wave background in GDT, we obtain a more general solution that explains the interaction dynamics of positons, dark positons and b-p. By choosing plane wave as the seed solution for both the components of CH equation we come across the second order b-p with four petaled b-p collides to form large amplitude b-p in one component and in the other component second order b-p with different amplitudes transform into four-petaled second order b-p.  On the other hand while choosing one component as the plane wave background and limiting the other component to zero, we visualize in one component, the second order degenerate bright soliton is reflected back by the second order b-p and in the other component second order degenerate dark soliton is reflected back by the second order b-p.  
 
	\par We present our work in the following manner. In Section 2, we describe the GDT method applicable for CH equation.  We then divide our presentation into two parts. In the first part, we explain the method of constructing positon solutions and hybrid solutions (consists of both solitons and positons) for the CH equation on zero background (Section 3).  In this section, the derivation of hybrid solutions are also explained.  In Section 4, we present the positon solutions on plane wave background. Further, we study the interaction dynamics of positons and b-ps. Finally, we summarize all the observed results in Section 5. 
	
	\section{Generalized Darboux transformation}
	Painlev\'{e} analysis \cite{ds-ch}, Hirota bilinear method \cite{HB,nld4}, B\"{a}cklund transformation \cite{BL} and Darboux transformation (DT) are some of the methods used to obtain exact solutions of nonlinear evolutionary equations. Among these, the DT method has been proved very effective in finding various kinds of localized solutions including multi-solitons, rogue wave, breathers and kinks \cite{soliton,dissipative,vp}.  In the conventional Darboux transformation (DT) method, iterative procedure will be adapted to generate distinct solutions with different spectral parameters. It has been demonstrated that the positon solutions can be constructed through GDT method by expressing the  eigenfunctions in Taylor expansion with the same spectral parameter \cite{m2,Matveevbook}.
  
 \par We derive the positon and b-p solutions of the coupled Hirota Eq.\eqref{e1} with the help of GDT method. The Lax pair of Eq.\eqref{e1} takes the form \cite{DT-che}
	\begin{subequations}
		\begin{eqnarray}
			\Phi_x &= &J \Phi,\quad \quad J = \lambda J_0 + J_1,\\
			\Phi_t& =& G \Phi, \quad \quad G= \lambda^3 G_0 + \lambda^2 G_1 + \lambda G_2 + G_3,
		\end{eqnarray}\label{lp}
		
		where 
		\begin{equation}
			J_0 = \dfrac{1}{12 \alpha} \begin{pmatrix}
				-2i & 0&0 \\0 &i & 0 \\0&0& i 
			\end{pmatrix},  J_1 = \begin{pmatrix}
				0& -\psi & -\phi\\ \psi^* & 0 & 0 \\ \phi^* & 0& 0
			\end{pmatrix}, \;G_0 = \dfrac{1}{16 \alpha} J_0,\; G_1 = \dfrac{1}{8 \alpha } J_0 + \dfrac{1}{16 \alpha } J_1,  \notag 
		\end{equation}
		\begin{eqnarray}
			G_ 2 &=& \dfrac{1}{4} \begin{pmatrix}
				i (|\psi|^2 + |\phi|^2) & - \dfrac{\psi}{ 2 \alpha}- i \psi_x& - \dfrac{\phi}{2 \alpha} - i \phi_x\\\\ \dfrac{\psi^*}{2 \alpha} - i \psi^* _x & - i |\psi|^2 & - i \phi \psi^*\\\\ \dfrac{\phi^*}{2 \alpha} - i \phi^*_x & - i \psi \phi^* & - i |\phi|^2
			\end{pmatrix},\notag \\ G_3 &= & \begin{pmatrix}
				\alpha 	(b_1 + b_2) + \dfrac{i}{2}( |\psi|^2 + |\phi|^2)
				& \alpha b_3 - \dfrac{i}{2} \psi_x & \alpha b_4 - \dfrac{i}{2} \phi_x \\ \\- \alpha b^*_3 - \dfrac{i}{2} \psi^*_x & -\alpha b_1 - \dfrac{i}{2} |\psi|^2 & \alpha b_5 - \dfrac{i}{2} \phi \psi^* \\\\
				
				- \alpha b^*_4 - \dfrac{i}{2} \phi^*_x & - \alpha b^*_5 - \dfrac{i}{2}\psi \phi^* & - \alpha b_2 - \dfrac{i}{2}|\phi|^2
			\end{pmatrix},\notag \\
		\end{eqnarray}
		with 
		\begin{eqnarray}
			b_1& =& \psi \psi^*_x - \psi^* \psi_x, \quad b_2 = \phi \phi^*_x - \phi^* \phi_x,\quad b_3 = \psi_{xx}+ 2 \alpha \psi, \quad \notag \\ b_4 &=&\phi_{xx}+ 2 \alpha \phi,\quad b_5 = \psi^* \phi_x - \phi \psi_x^*.  
		\end{eqnarray}
	\end{subequations}
	Here, $\Phi(x,t) = (u(x,t),v (x,t),w (x,t))^T$ are vector eigenfunctions, $\lambda$ is a complex spectral parameter and asterisk $(^*)$ represents its complex conjugate. Using the zero curvature condition, $J_t - G_x +[J,G] =0$, one can directly get Eq.\eqref{e1}. 
	\par In Ref. \cite{DT-che}, using GDT method, localized wave solutions on the plane background has been studied for Eq. \eqref{e1}. From the solutions of Lax pair (Eq.\eqref{lp}) with $\lambda_k$, $k=1,2....N$, as the spectral parameter, one can present the $N$-fold solution formula through the DT method \cite{DT-che} for the Eq.\eqref{e1} in the form 
	\begin{subequations}\label{dt}
		\begin{equation}
			\psi_N = \psi_0 + \dfrac{i}{4 \alpha} \dfrac{W_{2N}}{W_{1N}}, \qquad \phi_N = \phi_0 + \dfrac{i}{4 \alpha} \dfrac{W_{3N}}{W_{1N}},
		\end{equation}
		where $\psi_0, \phi_0$ are the seed solutions with
		\begin{equation}
			W_{1N} = \begin{vmatrix}
				u_1 \lambda_1^{N-1}& v_1 \lambda_1^{N-1}&w_1 \lambda_1^{N-1}& u_1 \lambda_1^{N-2}& \cdots& u_1 & v_1 & w_1 \\\\
				-v_1^* \lambda_1^{*N-1} &u_1^* \lambda_1^{*N-1}&0&-v_1^* \lambda_1^{*N-2}&\cdots &-v_1^*& u_1^*&0\\\\
				-w_1^* \lambda_1^{*N-1}& 0& u_1^* \lambda_1^{*N-1} & -w_1^* \lambda_1^{*N-2}& \cdots & -w_1^*& 0 &  u_1^*\\\\
				u_2 \lambda_2^{N-1}& v_2 \lambda_2^{N-1}&w_2 \lambda_2^{N-1}& u_2 \lambda_2^{N-2}& \cdots& u_2 & v_2 & w_2 \\ \vdots &  \vdots & \vdots & \vdots & \vdots & \vdots & \vdots & \vdots \\
				-w_N^* \lambda_N^{*N-1}& 0& u_N^* {\lambda_N^{*}}^{N-1} & -w_N^* {\lambda_N^{*}}^{N-2}& \cdots &- w_N^*& 0 &  u_N^*	
			\end{vmatrix},\label{w1}
		\end{equation}
		
		\begin{equation}
			W_{2N} = \begin{vmatrix}
				u_1 \lambda_1^{N-1}&u_1 \lambda_1^{N} &w_1 \lambda_1^{N-1}& u_1 \lambda_1^{N-2}& \cdots& u_1 & v_1 & w_1 \\\\
				-v_1^* \lambda_1^{*N-1} &-v_1^* \lambda_1^{*N}&0&-v_1^* \lambda_1^{*N-2}&\cdots &-v_1^*& u_1^*&0\\\\
				-w_1^* \lambda_1^{*N-1}&-	w_1^* \lambda_1^{*N} & u_1^* \lambda_1^{*N-1} &- w_1^* \lambda_1^{*N-2}& \cdots & -w_1^*& 0 & u_1^*\\\\
				u_2 \lambda_2^{N-1}&u_2 \lambda_2^{N}&w_2 \lambda_2^{N-1}& u_2 \lambda_2^{N-2}& \cdots& u_2 & v_2 & w_2 \\ \vdots &  \vdots & \vdots & \vdots & \vdots & \vdots & \vdots & \vdots \\
				-w_N^* \lambda_N^{*N-1}&- w_N^* \lambda_N^{*N}& u_N^* \lambda_N^{*N-1} &- w_N^* \lambda_N^{*N-2}& \cdots &- w_N^*& 0 & u_N^*	
			\end{vmatrix},\label{w2}
		\end{equation}
		and
		\begin{equation}
			W_{3N} = \begin{vmatrix}
				u_1 \lambda_1^{N-1}& v_1 \lambda_1^{N-1}&u_1 \lambda_1^{N}& u_1 \lambda_1^{N-2}& \cdots& u_1 & v_1 & w_1 \\\\
				-v_1^* \lambda_1^{*N-1} &u_1^* \lambda_1^{*N-1}&	-v_1^* \lambda_1^{*N}&-v_1^* \lambda_1^{*N-2}&\cdots &-v_1^*& u_1^*&0\\\\
				-w_1^* \lambda_1^{*N-1}& 0&	-w_1^* \lambda_1^{*N} & -w_1^* \lambda_1^{*N-2}& \cdots &- w_1^*& 0 & u_1^*\\\\
				u_2 \lambda_2^{N-1}& v_2 \lambda_2^{N-1}&u_2 \lambda_2^{N}& u_2 \lambda_2^{N-2}& \cdots& u_2 & v_2 & w_2 \\ \vdots &  \vdots & \vdots & \vdots & \vdots & \vdots & \vdots & \vdots \\
				-w_N^* \lambda_N^{*N-1}& 0&-w_N^* \lambda_N^{*N} &- w_N^* \lambda_N^{*N-2}& \cdots & -w_N^*& 0 &  u_N^*		
			\end{vmatrix}.\label{w3}
		\end{equation}
	\end{subequations}
	Here, $u_k, v_k$ and $w_k$ are the complex eigenfunctions with spectral parameter $\lambda_k$, $k=1,2...,N$. Substituting $N=2,3,...$ in Eq.\eqref{dt} with appropriate seed solutions, we can generate higher order localized wave  solutions including solitons, breathers and rogue waves.

	\section{Solutions of CH equation on zero background}
    In this section, we discuss the method of deriving positon solutions and positon-soliton solutions 
    of CH equation in detail.  For that we first solve the Lax pair Eq.\eqref{lp} with zero seed solution with the spectral parameter $\lambda= \lambda_k$ and obtain eigenfunctions in the form
	\begin{subequations}
		\begin{eqnarray}
			u_k(x,t)&=& s_{1k}~\exp\left[- \dfrac{i \lambda_z}{6 \alpha} x -\dfrac{i \lambda_k^2}{48 \alpha^2}\left( 1+ \dfrac{\lambda_k}{2}\right)t\right],\\	
			v_k(x,t)&=& s_{2k}~\exp\left[ \dfrac{i \lambda_k}{12 \alpha} x +\dfrac{i \lambda_k^2}{96 \alpha^2}\left( 1+ \dfrac{\lambda_k}{2}\right)t\right],\\
			w_k(x,t)&=& s_{3k}~\exp\left[ \dfrac{i \lambda_k}{12 \alpha} x +\dfrac{i \lambda_k^2}{96 \alpha^2}\left( 1+ \dfrac{\lambda_k}{2}\right)t\right],		
		\end{eqnarray}\label{ef1}
	\end{subequations}
where $k=1,2,3,...N$, $s_{1k}$, $s_{2k}$ and $s_{3k}$ are real arbitrary parameters. 
 These eigenfunctions can be used to derive soliton, positon and positon-soliton solutions of CH equation.  Upon substituting the above eigenfunctions as it is in the determinants (\ref{dt}) one can obtain N-soliton solutions of CH equation. In the next two subsections we describe the method of deriving positon and positon-soliton solutions for the Eq. \eqref{1}.
 
 \subsection{Positon solutions}
 To derive positon solutions, we restrict all the spectral parameters into $\lambda_k = \lambda_1+\epsilon$, $k=2,3,...N$, and expand the eigenfunctions in Taylor series at $\epsilon$, where $\epsilon$ is a real parameter. While doing so, we also amend the determinants of Eq.\eqref{dt} in the following  form \cite{w1,w2}
	\begin{eqnarray}
		\lim_{\epsilon\rightarrow0}	W_l[N]=\left|\dfrac{\partial^{r_i-1}}{\partial\epsilon^{r_i-1}}(\tilde{W}_{lN})\right|_{3N\times3N},\label{dt1}
	\end{eqnarray}
	where $\tilde{W}_{lN}=(W_{lN})_{ij}(\lambda_1+\epsilon)$, $l=1,2,3$, with  $r_i=\left[\dfrac{i+2}{3}\right]$, $[i]$ is the floor function. Here, $i$ and $j$ denotes the row and column of the matrix, respectively. 
	\begin{figure}[ht!]
		\centering
		\includegraphics[width=1\linewidth]{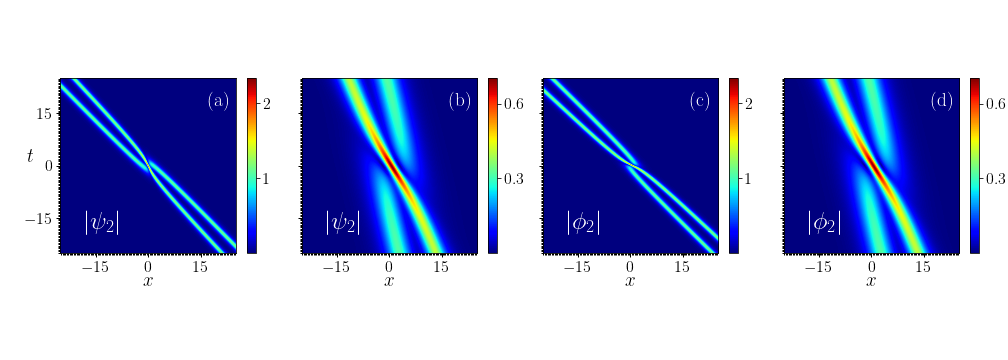}
		\caption{second order positon contour plots of $|\psi_2|$ and $|\phi_2|$ components with the parameter values $s_{12}, s_{22}, s_{32} =1$, $\lambda_1 = -1 + 2i$ with (a), (c) for $\alpha = 0.3$ and (b), (d) for $\alpha = 1$.}\label{1}
	\end{figure}
	\par For example, to capture the second order positon solution, we take $N=2$  and the seed solutions as $\psi_0 = \phi_0 = 0$ in Eq.\eqref{dt1}. Now using Taylor series, we expand the eigenfunctions \eqref{ef1} with respect to $\epsilon$ by confining the spectral parameter $\lambda_2$ as $\lambda_2 = \lambda_1 + \epsilon$ and substitute the obtained eigenfunctions in the GDT modified formula \eqref{dt1} and obtain the exact expression of second order positon solution as given below
	\begin{subequations}\label{2ps}
		\begin{equation}
			\psi_2(x,t) =\dfrac{D_1}{D_2},\quad  \phi_2(x,t) = \dfrac{D_3}{D_2},	
		\end{equation}  
		where
		\begin{eqnarray}
			D_1 &=&	16 s_{12} s_{22} \alpha (\lambda_1 - \lambda_1^*) (e^{\frac{
					i \lambda_1 (16 x \alpha + t \lambda_1 (2 + \lambda_1))}{
					64 \alpha^2}} (s_{22}^2 + s_{32}^2) (128 i \alpha^2  + 
			16 x \alpha (\lambda_1 - \lambda_1^*) \notag \\&&+ 
			t \lambda_1 (4 + 3 \lambda_1) (\lambda_1 - \lambda_1^*)) + e^{\frac{
					i \lambda_1^* (16 x \alpha + t \lambda_1^* (2 + \lambda_1^*))}{
					64 \alpha^2}} 
			s_{12}^2 (128 i \alpha^2 - 
			16 x \alpha (\lambda_1 - \lambda_1^*)\notag \\&& - 
			t (\lambda_1 - \lambda_1^*) \lambda_1^* (4 + 
			3 \lambda_1^*))),\\
			D_2 &=&4096 e^{\frac{
					i \lambda_1^* (16 x \alpha + t \lambda_1^* (2 + \lambda_1^*))}{
					32 \alpha^2}} s_{12}^4 \alpha^4 + 
			4096 e^{\frac{
					i \lambda_1 (16 x \alpha + t \lambda_1 (2 + \lambda_1))}{
					64 \alpha^2}}  (s_{22}^2 + s_{32}^2)^2 \alpha^4 \notag \\&& + 
			e^{\frac{i (16 x \alpha (\lambda_1 + \lambda_1^*) + 
					t (2 \lambda_1^2 + \lambda_1^3 + \lambda_1^{*2} (2 + \lambda_1^*)))}{64 \alpha^2}}
			s_{12}^2 (s_{22}^2 + s_{32}^2) (8192 \alpha^4 - 
			256 x^2 \alpha^2 (\lambda_1 - \lambda_1^*)^2 \notag \\&&- 
			t^2 \lambda_1 (4 + 3 \lambda_1) (\lambda_1 - \lambda_1^*)^2 \lambda_1^* (4 + 
			3 \lambda_1^*) - 
			16 t x \alpha (\lambda_1 - \lambda_1^*)^2   (4 \lambda_1 + 
			3 \lambda_1^2 \notag \\&& + \lambda_1^* (4 + 3 \lambda_1^*))),\\
			D_3&=& 16 s_{12} s_{32} \alpha (\lambda_1 - \lambda_1^*) (e^{\frac{
					i \lambda_1 (16 x \alpha + t \lambda_1 (2 + \lambda_1))}{
					64 \alpha^2}} (s_{22}^2 + s_{32}^2) (128 i \alpha^2 + 
			16 x \alpha (\lambda_1 - \lambda_1^*) \notag \\&& + 
			t \lambda_1 (4 + 3 \lambda_1) (\lambda_1 - \lambda_1^*)) + e^{\frac{
					i \lambda_1^* (16 x \alpha + t \lambda_1^* (2 + \lambda_1^*))}{
					64 \alpha^2}} 
			s_{12}^2 (128 i \alpha^2 - 
			16 x \alpha (\lambda_1 - \lambda_1^*) \notag \\&& - 
			t (\lambda_1 - \lambda_1^*) \lambda_1^* (4 + 
			3 \lambda_1^*))). 
		\end{eqnarray}
	\end{subequations}
	\par The second order positon plots of $|\psi_2|$ and $|\phi_2|$ are presented in Fig.~\ref{1}. We examine how the higher order nonlinear term $\alpha$ influences the positon solutions. Strengthening the system paramater ($\alpha$), the width of the second order positons gets enhanced and the distance between the two waves gets widened in both the components, see Fig. \ref{1}(a)-(d). One can notice that the second order positon has two waves of equal amplitude travelling with same velocity. \par
\begin{figure}[ht!]
		\centering
		\includegraphics[width=1\linewidth]{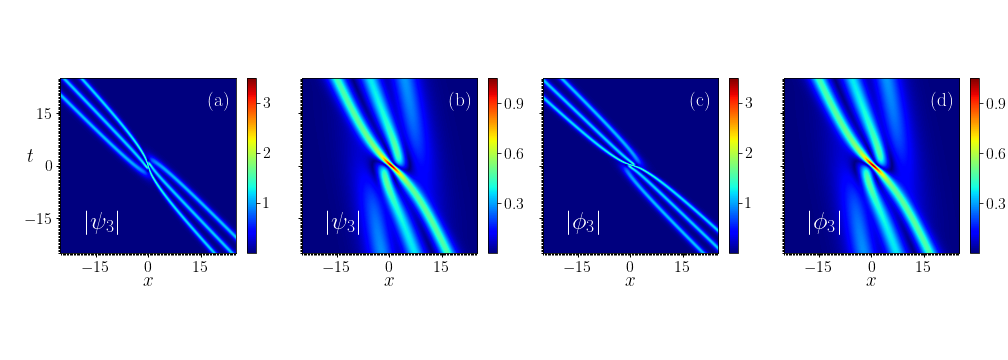}
		\caption{Third order positon contour plots of $|\psi_3|$ and $|\phi_3|$ components with the parameter values  $s_{13}, s_{23}, s_{33} =1$ $\lambda_1 = -1 + 2i$ with (a), (c) for $\alpha = 0.3$ and (b), (d) for $\alpha = 1$.}\label{2}
	\end{figure}

\begin{figure}[ht!]
		\centering
		\includegraphics[ width=1\linewidth]{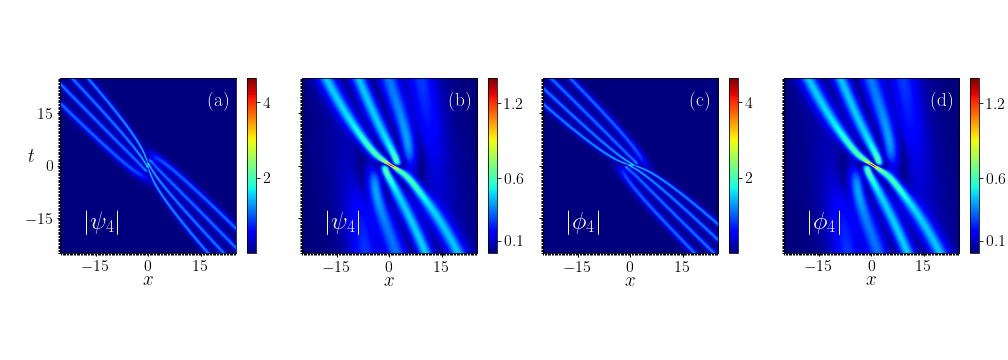}
		\caption{Fourth order positon contour plots of $|\psi_4|$ and $|\phi_4|$ components with the parameter values  $s_{14}, s_{24}, s_{34} =1$ $\lambda_1 = -1 + 2i$ with (a), (c) for $\alpha = 0.3$ and (b), (d) for $\alpha = 1$.}\label{3}
	\end{figure}

 The cases $N=3$ and $4$ in \eqref{dt1} with the same spectral paramater ($\lambda_1$) generates the third and fourth order smooth positon solutions. These two  higher order smooth positons also exhibit the same behaviour as we observed in the second order positon case which is evident from Figs.~\ref{2} and \ref{3}.

	\subsubsection{Asymptotic behaviour of second order positon solution}
	Next, we investigate the asymptotic nature of the second order positon solution for Eq. \eqref{e1}. As one can see in Eq. \eqref{2ps}, in the second order positon solution, the variable $t$ exist in both polynomial and exponential form, hence it becomes zero when $t\rightarrow\pm\infty$. Therefore, we cannot apply the conventional asymptotic analysis to the positon solution. So we adapt the procedure given in \cite{w2,kdv,Hirota2}. In this procedure, we equate the absolute maximal amplitude of one soliton with the solution of shifted second order positon ($\psi_2$ and $\phi_2$) in the asymptotic limits and capture their behaviour. 
	\par First, we determine the highest amplitude of one soliton by substituting $N=1$ in the solution formulae \eqref{dt}, with $\lambda_1 = -c+i d$. We can obtain one soliton solution for the component $\psi_1(x,t)$ in the form
{\begin{equation}
		\psi_1(x,t) = \dfrac{- d s_{11} s_{21} \exp \left[{\frac{(c-id)^2 ( i(-1+c)+d)t+4(ic+d)x\alpha}{8 \alpha^2}}\right]}{\alpha \left(s_{21}^2 + s_{31}^2 + s_{11}^2 \exp \left[{\frac{d( 4 x \alpha-(2c-3c^2+d^2)t )}{4  \alpha^2}}\right]\right)},
	\end{equation}}
	where $c$ and $d$ are real parameters. The absolute value of $\psi_1(x,t)$ reads
	\begin{equation}
		|\psi_1(x,t)| = \dfrac{d s_{11} s_{21}  \exp \left[{\frac{d(4 x \alpha-(2c-3c^2+d^2)t)}{8 \alpha^2}}\right]}{\alpha \left(s_{21}^2 + s_{31}^2 + s_{11}^2 \exp \left[{\frac{d(4 x \alpha-(2c-3c^2+d^2)t)}{4 \alpha^2}}\right]\right)}.
	\end{equation}
	The absolute maximum  of one soliton can be evaluated from the above expression which inturn reads
	\begin{equation}
		|\psi_1(\hat{x},t)|= \dfrac{ s_{21} d}{2  \alpha \sqrt{s_{21}^2+s_{31}^2}}.\label{1a}
	\end{equation}

\par Similarly, the one soliton solution of the $\phi_1$ component takes the form 
	\begin{equation}
		\phi_1(x,t) = \dfrac{- d s_{11} s_{31} \exp \left[{\frac{(c-id)^2 ( i(-1+c)+d)t+4(ic+d)x\alpha}{8 \alpha^2}}\right]}{\alpha \left(s_{21}^2 + s_{31}^2 + s_{11}^2 \exp \left[{\frac{d( 4 x \alpha-(2c-3c^2+d^2)t )}{4  \alpha^2}}\right]\right)},
	\end{equation}
The absolute value of $\phi_1(x,t)$ with the maximum point $\hat{x}$ is given by 
	\begin{equation}
		|\phi_1(\hat{x},t)|= \dfrac{ s_{31} d}{2  \alpha \sqrt{s_{21}^2+s_{31}^2}},\label{a1}
	\end{equation}
 where the maximum point ($\hat{x}$) of $\psi_1(x,t)$ and $\phi_1(x,t)$ is identified as
	\begin{equation}
		\hat{x}=\dfrac{(2c-3c^2+d^2)t}{4 \alpha} + \dfrac{2 \alpha \log[\frac{\sqrt{s_{21}^2+s_{31}^2}}{s_{11}}]}{d}.	 
	\end{equation} 
\par By comparing the maximum amplitude of one soliton solution, \eqref{1a} and \eqref{a1},   with leading terms of $t$ in the absolute second order positon solution \eqref{2ps}, we can identify a time-dependent displacement function in the form 
	\begin{equation}
		\Delta(x,t) = \dfrac{2 \alpha}{d} \left(\log\left[d^2 \sqrt{(1- 3c)^2+d^2} \right]t - \log\left[ 2 \alpha^2 \right] \right). \label{delta}
	\end{equation}
\par In the above Eq. \eqref{delta}, we can notice that the time-dependent displacement function is directly proportional to the nonlinear parameter $\alpha$. In the second order positon solution one witnesses the presence of trignometric terms in the component $\psi_2(x,t)$ and $\phi_2(x,t)$. We denote the arguments of these trigonometric functions as $A(x,t)$. Since the internal oscillations are caused by various values of $A(x,t)$,  we fix this function $A(x,t)$ as a constant and determine the maximum amplitude of second order positon solution \cite{w2}. We replace $x$ by $ \hat{x}\pm \Delta(t)$ in the second order positon solution \eqref{2ps} with constant $A$. Now, picking up the leading order terms of $t$, we can identify the following expressions, namely
	\begin{subequations}\label{asy} \scriptsize
		\begin{equation}
			\lim_{t\rightarrow\pm \infty}
			\psi_2(\hat{x}+\Delta(t),t) =\mp \dfrac{d s_{21} \left(\left(d\cos[A]+(3c-1)\sin[A]\right) \mp i \left((1 - 3c)\cos[A] + d\sin[A] \right)\right)}{2 \alpha\sqrt{(s_{21}^2+s_{31}^2)((1-3c)^2 + d^2)}},		
		\end{equation}
		\\
		\begin{equation}
			\lim_{t\rightarrow\pm \infty}\psi_2(\hat{x}-\Delta(t),t) =\pm \dfrac{d s_{21} \left(\left(d\cos[A]+(1-3c)\sin[A]\right) \pm i \left(( 3c-1)\cos[A] + d\sin[A] \right)\right)}{2 \alpha\sqrt{(s_{21}^2+s_{31}^2)((1-3c)^2 + d^2)}},		
		\end{equation}
		\\\begin{equation}
			\lim_{t\rightarrow\pm \infty}
			\phi_2(\hat{x}+\Delta(t),t) =\mp \dfrac{d s_{31} \left(\left(d\cos[A]+(3c-1)\sin[A]\right) \mp i \left((1 - 3c)\cos[A] + d\sin[A] \right)\right)}{2 \alpha\sqrt{(s_{21}^2+s_{31}^2)((1-3c)^2 + d^2)}},		
		\end{equation}
		\\
		\begin{equation}
			\lim_{t\rightarrow\pm \infty}\phi_2(\hat{x}-\Delta(t),t) =\pm \dfrac{d s_{31} \left(\left(d\cos[A]+(1-3c)\sin[A]\right) \pm i \left(( 3c-1)\cos[A] + d\sin[A] \right)\right)}{2 \alpha\sqrt{(s_{21}^2+s_{31}^2)((1-3c)^2 + d^2)}},		
		\end{equation}
		where
		\begin{equation}
			A(x,t)=  \dfrac{4 c  \alpha x+(d^2 + c ((c-1) c - 3 d^2)) t}{8 \alpha^2}.
		\end{equation}
	\end{subequations}
From \eqref{asy}, we can capture the absolute maximum of the shifted second order positon solution as
	
	\begin{equation}
		|\psi_2(\hat{x}\pm \Delta(t),t)|= \dfrac{ s_{21} d}{2  \alpha \sqrt{s_{21}^2+s_{31}^2}},\quad |\phi_2(\hat{x}\pm \Delta(t),t)|=  \dfrac{ s_{31} d}{2  \alpha \sqrt{s_{21}^2+s_{31}^2}}.\label{2a}
	\end{equation}
	\par Upon comparing the Eqs. \eqref{1a}, \eqref{a1} and \eqref{2a}, we can conclude that the maximum amplitude of one soliton ($\psi_1$, $\phi_1$) and the shifted second order positon solution in both the components ($\psi_2$, $\phi_2$) are similar as soliton solutions. But the special feature of positon solutions is that it exhibits time-dependent displacement. Due to this property, the second order positon travel as a single component for smaller time scales and it will separate into the constituents of two `one solitons' over large time \cite{Hirota2}. This phase shift or displacement function (see Eq. \eqref{delta}) also depends on the nonlinear parameter $\alpha$. This is the reason for, while we increase the value of the parameter $\alpha$, the distance between second order positons increases. 
 \subsection{Hybrid solutions (Positon - Soliton solutions)}
	In this subsection, we unearth a hybrid solution which is a combination of soliton and positon solution of the CH Eq. \eqref{e1} and investigate the interactions between them.  We extract this hybrid solution by recasting the determinant \eqref{dt}.  In the following, we explain how it can be extracted from the determinants.\\
 \newline \textbf{Case 1: Interaction between second order positon and one soliton}\\
 \par To construct second order positon with one soliton solution, we consider $N=3$ in the DT formula (\ref{dt}) with seed solutions $\psi_0=0$ and $\phi_0=0$. The resultant solution can be represented in the form 
		\begin{equation}
			\psi_3 = \dfrac{i}{4 \alpha} \dfrac{W_{23}}{W_{13}}, \qquad \phi_3 = \dfrac{i}{4 \alpha} \dfrac{W_{33}}{W_{13}},
        \label{dt2}
		\end{equation}
 The explicit determinant expressions of $W_{23}$, $W_{31}$ and $W_{13}$ are given by
 \begin{subequations}\label{3s}
		\begin{eqnarray}
			{W}_{13} &=& \begin{vmatrix}
				u_1 \lambda_1^{2}& v_1 \lambda_1^{2}&w_1 \lambda_1^{2}& u_1 \lambda_1&v_1 \lambda_1 &w_1 \lambda_1& u_1 & v_1 & w_1 \\
				-v_1^* \lambda_1^{*2} &u_1^* \lambda_1^{*2}&0& -v_1^* \lambda_1^*&u_1^* \lambda_1^{*}&0&-v_1^*& u_1^*&0\\
				-w_1^* \lambda_1^{*2}& 0& u_1^* \lambda_1^{*2} & -w_1^* \lambda_1^{*}& 0&u_1^* \lambda_1^{*} & -w_1^*& 0 &  u_1^*\\
				u_{{2}} \lambda_{2}^{2}& v_{{2}} \lambda_{2}^{2}&w_{{2}} \lambda_{2}^{2}& u_{{2}} \lambda_{2}&v_{{2}} \lambda_{2}&w_{2} \lambda_{2}& u_{2} & v_{2} & w_{2} \\
				-v_{2}^* \lambda_{2}^{*2} &u_{2}^* \lambda_{2}^{*2}&0& -v_{2}^* \lambda_{2}^*&u_{2}^* \lambda_{2}^{*}&0&{-v_{2}^*}& {u_{2}^{*}}&0\\
				-w_{2}^* \lambda_{2}^{*2}& 0& u_{2}^* \lambda_{2}^{*2} & -w_{2}^* \lambda_{2}^{*}& 0&u_{2}^* \lambda_{2}^{*} & {-w_{2}^*}& 0 & { u_{2}^*}\\
				u_3 \lambda_3^{2}& v_3 \lambda_3^{2}&w_3 \lambda_3^{2}& u_3 \lambda_3&v_3 \lambda_3 &w_3 \lambda_3& u_3 & v_3 & w_3 \\
				-v_3^* \lambda_3^{*2} &u_3^* \lambda_3^{*2}&0& -v_3^* \lambda_3^*&u_3^* \lambda_3^{*}&0&-v_3^*& u_3^*&0\\
				-w_3^* \lambda_3^{*2}& 0& u_3^* \lambda_3^{*2} &- w_3^* \lambda_3^{*}& 0&u_3^* \lambda_3^{*} & -w_3^*& 0 &  u_3^*
			\end{vmatrix},\notag\\
   \end{eqnarray}
   \begin{eqnarray}
			{W}_{23} &=& \begin{vmatrix}
				u_1 \lambda_1^{2}& u_1 \lambda_1^{3}&w_1 \lambda_1^{2}& u_1 \lambda_1&v_1 \lambda_1 &w_1 \lambda_1& u_1 & v_1 & w_1 \\
				-v_1^* \lambda_1^{*2} &-v_1^* \lambda_1^{*3}&0& -v_1^* \lambda_1^*&u_1^* \lambda_1^{*}&0&-v_1^*& u_1^*&0\\
				-w_1^* \lambda_1^{*2}& -w_1^* \lambda_1^{*3}& u_1^* \lambda_1^{*2} &- w_1^* \lambda_1^{*}& 0&u_1^* \lambda_1^{*} & -w_1^*& 0 &  u_1^*\\
				u_{{2}} \lambda_{2}^{2}& u_{{2}} \lambda_{2}^{3}&w_{{2}} \lambda_{2}^{2}& u_{{2}} \lambda_{2}&v_{{2}} \lambda_{2}&w_{2} \lambda_{2}& u_{2} & v_{2} & w_{2} \\
				-v_{2}^* \lambda_{2}^{*2} &-v_{2}^* \lambda_{2}^{*3}&0& -v_{2}^* \lambda_{2}^*&u_{2}^* \lambda_{2}^{*}&0&{-v_{2}^*}& {u_{2}^{*}}&0\\
				-w_{2}^* \lambda_{2}^{*2}&-w_{2}^* \lambda_{2}^{*3} & u_{2}^* \lambda_{2}^{*3} & -w_{2}^* \lambda_{2}^{*}& 0&u_{2}^* \lambda_{2}^{*}& {-w_{2}^*}& 0 & { u_{2}^*}\\
				u_3 \lambda_3^{2}& u_3 \lambda_3^{3}&w_3 \lambda_3^{2}& u_3 \lambda_3&v_3 \lambda_3 &w_3 \lambda_3& u_3 & v_3 & w_3 \\
				-v_3^* \lambda_3^{*2} &-v_3^* \lambda_3^{*2}&0& -v_3^* \lambda_3^*&u_3^* \lambda_3^{*}&0&-v_3^*& u_3^*&0\\
				-w_3^* \lambda_3^{*2}&-w_3^* \lambda_3^{*3} & u_3^* \lambda_3^{*2} &- w_3^* \lambda_3^{*}& 0&u_3^* \lambda_3^{*} & -w_3^*& 0 &  u_3^*
			\end{vmatrix}, \notag \\ \end{eqnarray}
   \begin{eqnarray}
	{W}_{33}& =& \begin{vmatrix}
				u_1 \lambda_1^{2}& v_1 \lambda_1^{2}&u_1 \lambda_1^{3}& u_1 \lambda_1&v_1 \lambda_1 &w_1 \lambda_1& u_1 & v_1 & w_1 \\
				-v_1^* \lambda_1^{*2} &u_1^* \lambda_1^{*2}&-v_1^* \lambda_1^{*3} & -v_1^* \lambda_1^*&u_1^* \lambda_1^{*}&0&-v_1^*& u_1^*&0\\
				-w_1^* \lambda_1^{*2}& 0& -w_1^* \lambda_1^{*3} & w_1^* \lambda_1^{*}& 0&u_1^* \lambda_1^{*} & -w_1^*& 0 &  u_1^*\\
				u_{{2}} \lambda_{2}^{2}& v_{{2}} \lambda_{2}^{2}&u_{{2}} \lambda_{2}^{3}& u_{{2}} \lambda_{2}&v_{{2}} \lambda_{2}&w_{2} \lambda_{2}& u_{2} & v_{2} & w_{2} \\
				-v_{2}^* \lambda_{2}^{*2} &u_{2}^* \lambda_{2}^{*2}&-v_{2}^* \lambda_{2}^{*3}& -v_{2}^* \lambda_{2}^*&u_{2}^* \lambda_{2}^{*}&0&{-v_{2}^*}& {u_{2}^{*}}&0\\
			-w_{2}^* \lambda_{2}^{*2}& 0& -w_{2}^* \lambda_{2}^{*3} & -w_{2}^* \lambda_{2}^{*}& 0&u_{2}^* \lambda_{2}^{*} & {-w_{2}^*}& 0 & { u_{2}^*}\\
				u_3 \lambda_3^{2}& v_3 \lambda_3^{2}&u_3 \lambda_3^{3}& u_3 \lambda_3&v_3 \lambda_3 &w_3 \lambda_3& u_3 & v_3 & w_3 \\
				-v_3^* \lambda_3^{*2} &u_3^* \lambda_3^{*2}&-v_3^* \lambda_3^{*3}& -v_3^* \lambda_3^*&u_3^* \lambda_3^{*}&0&-v_3^*& u_3^*&0\\
				-w_3^* \lambda_3^{*2}& 0& -w_3^* \lambda_3^{*3} &- w_3^* \lambda_3^{*}& 0&u_3^* \lambda_3^{*} & -w_3^*& 0 &  u_3^*
			\end{vmatrix}. \notag \\
		\end{eqnarray}
		\end{subequations}
The above determinant expressions comprise of three eigenvalues namely, $\lambda_1$, $\lambda_2$ and $\lambda_3$.  As we are interested in deriving second order positon and one soliton solution, we limit the spectral parameter $\lambda_2 \rightarrow \lambda_1 + \epsilon$ and differentiate only the eigenfuction that corresponds to spectral parameter $\lambda_2$ with respect to $\epsilon$. Since the parameter $\lambda_3$ and its corresponding eigenfuctions remain unchanged it gives rise to the evolution of soliton.  However the validity of the solution can be verified using $Mathematica$. The explicit expression of this second order positon - one soliton solution is very lengthy, in the following, we express only the form of determinants. 
	\begin{subequations} \label{hdt} \scriptsize
		\begin{equation}
			\lim_{\epsilon\rightarrow0}\widehat{W}_{13} = \begin{vmatrix}
				u_1 \lambda_1^{2}& v_1 \lambda_1^{2}&w_1 \lambda_1^{2}& u_1 \lambda_1&v_1 \lambda_1 &w_1 \lambda_1& u_1 & v_1 & w_1 \\
				-v_1^* \lambda_1^{*2} &u_1^* \lambda_1^{*2}&0& -v_1^* \lambda_1^*&u_1^* \lambda_1^{*}&0&-v_1^*& u_1^*&0\\
				-w_1^* \lambda_1^{*2}& 0& u_1^* \lambda_1^{*2} & -w_1^* \lambda_1^{*}& 0&u_1^* \lambda_1^{*} & -w_1^*& 0 &  u_1^*\\
				[u_{{1\epsilon}} \lambda_{1\epsilon}^{2}]'& [v_{{1\epsilon}} \lambda_{1\epsilon}^{2}]'&[w_{{1\epsilon}} \lambda_{1\epsilon}^{2}]'& [u_{{1\epsilon}} \lambda_{1\epsilon}]'&[v_{{1\epsilon}} \lambda_{1\epsilon}]'&[w_{1\epsilon} \lambda_{1\epsilon}]'& u_{1\epsilon}' & v_{1\epsilon}' & w_{1\epsilon}' \\
				[-v_{1\epsilon}^* \lambda_{1\epsilon}^{*2}]' &[u_{1\epsilon}^* \lambda_{1\epsilon}^{*2}]'&0& [-v_{1\epsilon}^* \lambda_{1\epsilon}^*]'&[u_{1\epsilon}^* \lambda_{1\epsilon}^{*}]'&0&{-v_{1\epsilon}^*}'& {u_{1\epsilon}^{*}}'&0\\
				[-w_{1\epsilon}^* \lambda_{1\epsilon}^{*2}]'& 0& [u_{1\epsilon}^* \lambda_{1\epsilon}^{*2}]' & [-w_{1\epsilon}^* \lambda_{1\epsilon}^{*}]'& 0&[u_{1\epsilon}^* \lambda_{1\epsilon}^{*} ]'& {-w_{1\epsilon}^*}'& 0 & { u_{1\epsilon}^*}'\\
				u_3 \lambda_3^{2}& v_3 \lambda_3^{2}&w_3 \lambda_3^{2}& u_3 \lambda_3&v_3 \lambda_3 &w_3 \lambda_3& u_3 & v_3 & w_3 \\
				-v_3^* \lambda_3^{*2} &u_3^* \lambda_3^{*2}&0& -v_3^* \lambda_3^*&u_3^* \lambda_3^{*}&0&-v_3^*& u_3^*&0\\
				-w_3^* \lambda_3^{*2}& 0& u_3^* \lambda_3^{*2} &- w_3^* \lambda_3^{*}& 0&u_3^* \lambda_3^{*} & -w_3^*& 0 &  u_3^*
			\end{vmatrix},
		\end{equation}\\
		\begin{equation}
			\lim_{\epsilon\rightarrow0}	\widehat{W}_{23} = \begin{vmatrix}
				u_1 \lambda_1^{2}& u_1 \lambda_1^{3}&w_1 \lambda_1^{2}& u_1 \lambda_1&v_1 \lambda_1 &w_1 \lambda_1& u_1 & v_1 & w_1 \\
				-v_1^* \lambda_1^{*2} &-v_1^* \lambda_1^{*3}&0& -v_1^* \lambda_1^*&u_1^* \lambda_1^{*}&0&-v_1^*& u_1^*&0\\
				-w_1^* \lambda_1^{*2}& -w_1^* \lambda_1^{*3}& u_1^* \lambda_1^{*2} &- w_1^* \lambda_1^{*}& 0&u_1^* \lambda_1^{*} & -w_1^*& 0 &  u_1^*\\
				[u_{{1\epsilon}} \lambda_{1\epsilon}^{2}]'& [u_{{1\epsilon}} \lambda_{1\epsilon}^{3}]'&[w_{{1\epsilon}} \lambda_{1\epsilon}^{2}]'& [u_{{1\epsilon}} \lambda_{1\epsilon}]'&[v_{{1\epsilon}} \lambda_{1\epsilon}]'&[w_{1\epsilon} \lambda_{1\epsilon}]'& u_{1\epsilon}' & v_{1\epsilon}' & w_{1\epsilon}' \\
				[-v_{1\epsilon}^* \lambda_{1\epsilon}^{*2}]' &[-v_{1\epsilon}^* \lambda_{1\epsilon}^{*3}]'&0& [-v_{1\epsilon}^* \lambda_{1\epsilon}^*]'&[u_{1\epsilon}^* \lambda_{1\epsilon}^{*}]'&0&{-v_{1\epsilon}^*}'& {u_{1\epsilon}^{*}}'&0\\
				[-w_{1\epsilon}^* \lambda_{1\epsilon}^{*2}]'&[-w_{1\epsilon}^* \lambda_{1\epsilon}^{*3}]' & [u_{1\epsilon}^* \lambda_{1\epsilon}^{*3}]' & [-w_{1\epsilon}^* \lambda_{1\epsilon}^{*}]'& 0&[u_{1\epsilon}^* \lambda_{1\epsilon}^{*} ]'& {-w_{1\epsilon}^*}'& 0 & { u_{1\epsilon}^*}'\\
				u_3 \lambda_3^{2}& u_3 \lambda_3^{3}&w_3 \lambda_3^{2}& u_3 \lambda_3&v_3 \lambda_3 &w_3 \lambda_3& u_3 & v_3 & w_3 \\
				-v_3^* \lambda_3^{*2} &-v_3^* \lambda_3^{*2}&0& -v_3^* \lambda_3^*&u_3^* \lambda_3^{*}&0&-v_3^*& u_3^*&0\\
				-w_3^* \lambda_3^{*2}&-w_3^* \lambda_3^{*3} & u_3^* \lambda_3^{*2} &- w_3^* \lambda_3^{*}& 0&u_3^* \lambda_3^{*} & -w_3^*& 0 &  u_3^*
			\end{vmatrix},
		\end{equation}\\
		\begin{equation}
			\lim_{\epsilon\rightarrow0}\widehat{W}_{33} = \begin{vmatrix}
				u_1 \lambda_1^{2}& v_1 \lambda_1^{2}&u_1 \lambda_1^{3}& u_1 \lambda_1&v_1 \lambda_1 &w_1 \lambda_1& u_1 & v_1 & w_1 \\
				-v_1^* \lambda_1^{*2} &u_1^* \lambda_1^{*2}&-v_1^* \lambda_1^{*3} & -v_1^* \lambda_1^*&u_1^* \lambda_1^{*}&0&-v_1^*& u_1^*&0\\
				-w_1^* \lambda_1^{*2}& 0& -w_1^* \lambda_1^{*3} & w_1^* \lambda_1^{*}& 0&u_1^* \lambda_1^{*} & -w_1^*& 0 &  u_1^*\\
				[u_{{1\epsilon}} \lambda_{1\epsilon}^{2}]'& [v_{{1\epsilon}} \lambda_{1\epsilon}^{2}]'&[u_{{1\epsilon}} \lambda_{1\epsilon}^{3}]'& [u_{{1\epsilon}} \lambda_{1\epsilon}]'&[v_{{1\epsilon}} \lambda_{1\epsilon}]'&[w_{1\epsilon} \lambda_{1\epsilon}]'& u_{1\epsilon}' & v_{1\epsilon}' & w_{1\epsilon}' \\
				[-v_{1\epsilon}^* \lambda_{1\epsilon}^{*2}]' &[u_{1\epsilon}^* \lambda_{1\epsilon}^{*2}]'&[-v_{1\epsilon}^* \lambda_{1\epsilon}^{*3}]'& [-v_{1\epsilon}^* \lambda_{1\epsilon}^*]'&[u_{1\epsilon}^* \lambda_{1\epsilon}^{*}]'&0&{-v_{1\epsilon}^*}'& {u_{1\epsilon}^{*}}'&0\\
				[-w_{1\epsilon}^* \lambda_{1\epsilon}^{*2}]'& 0& [-w_{1\epsilon}^* \lambda_{1\epsilon}^{*3}]' & [-w_{1\epsilon}^* \lambda_{1\epsilon}^{*}]'& 0&[u_{1\epsilon}^* \lambda_{1\epsilon}^{*} ]'& {-w_{1\epsilon}^*}'& 0 & { u_{1\epsilon}^*}'\\
				u_3 \lambda_3^{2}& v_3 \lambda_3^{2}&u_3 \lambda_3^{3}& u_3 \lambda_3&v_3 \lambda_3 &w_3 \lambda_3& u_3 & v_3 & w_3 \\
				-v_3^* \lambda_3^{*2} &u_3^* \lambda_3^{*2}&-v_3^* \lambda_3^{*3}& -v_3^* \lambda_3^*&u_3^* \lambda_3^{*}&0&-v_3^*& u_3^*&0\\
				-w_3^* \lambda_3^{*2}& 0& -w_3^* \lambda_3^{*3} &- w_3^* \lambda_3^{*}& 0&u_3^* \lambda_3^{*} & -w_3^*& 0 &  u_3^*
			\end{vmatrix},
		\end{equation}
		\end{subequations}
		  where
  \begin{eqnarray}
u_{1\epsilon}(x,t)&=& s_{12}~\exp\left[- \dfrac{i (\lambda_1+\epsilon)}{6 \alpha} x -\dfrac{i (\lambda_1+\epsilon)^2}{48 \alpha^2}\left( 1+ \dfrac{(\lambda_1+\epsilon)}{2}\right)t\right],\notag\\	
v_{1\epsilon}(x,t)&=& s_{22}~\exp\left[ \dfrac{i (\lambda_1+\epsilon)}{12 \alpha} x +\dfrac{i (\lambda_1+\epsilon)^2}{96 \alpha^2}\left( 1+ \dfrac{(\lambda_1+\epsilon)}{2}\right)t\right],\notag\\
w_{1\epsilon}(x,t)&=& s_{32}~\exp\left[ \dfrac{i (\lambda_1+\epsilon)}{12 \alpha} x +\dfrac{i (\lambda_1+\epsilon)^2}{96 \alpha^2}\left( 1+ \dfrac{(\lambda_1+\epsilon)}{2}\right)t\right],
\end{eqnarray}\label{ef}
with $\lambda_{1\epsilon}= \lambda_1 + \epsilon$. In Eq. \eqref{hdt}, prime ($'$) represents the differentiation with respect to $\epsilon$. In the above, the expressions for $u_1$, $v_1$ and $w_1$ are obtained from \eqref{ef1} by taking $k=1$.  Similarly, $u_3$, $v_3$ and $w_3$ are taken from \eqref{ef1} with $k=3$.  Next we study the interaction dynamics between positons and soliton. 
\par
\begin{figure}[ht!]
		\centering
		\includegraphics[width=1\linewidth]{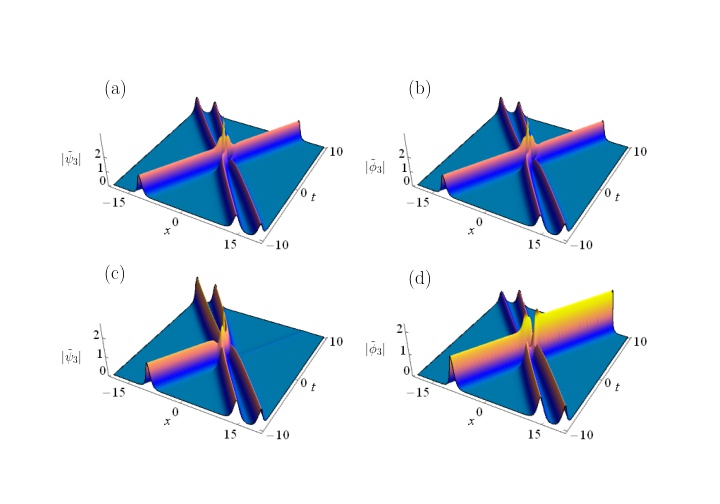}
		\caption{Interaction between second order positon and one soliton with parameter values $\lambda_1 = 13/20 +3/4 i $, $\lambda_3 =-3/4+i $ and $\alpha = 1/8 $. (a),(b)$s_{13} =s_{23} =s_{33} =1$; (c),(d) $s_{33} = 100$.} \label{4}
	\end{figure}

 The elastic and inelastic interaction between one soliton and second order positon are given in Fig. \ref{4}. In Fig. \ref{4}(a) and \ref{4}(b), one can visualize the elastic interaction that occur between second order positon and one soliton in the $|\tilde\psi_3|$ and $|\tilde\phi_3|$ components. Figures \ref{4}(c) and \ref{4}(d) reveal the inelastic interaction that occur between second order positon and one soliton for a higher value of arbitrary parameters ($s_{33}$ or $s_{23}$). Strengthening the value of the parameter $s_{33}$, causes the amplitude of one soliton to decrease and tend to disappear and it causes enhancement in the amplitude of the second order positon in $|\tilde\psi_3|$ component, see Figs. \ref{4}(c). In addition, the amplitudes of one soliton increases and one of the amplitudes of the second order positon is suppressed in  $|\tilde\phi_3|$ component, see Fig. \ref{4}(d) and it becomes vice versa while enhancing the parameter $s_{23}$. Here, the positon exhibits energy redistribution while colliding with the soliton.

	\par
\begin{figure}[ht!]
		\centering
		\includegraphics[width=1\linewidth]{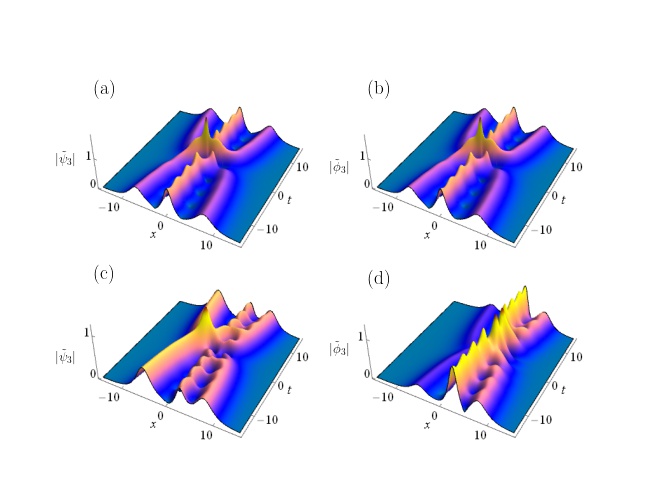}
		\caption{Bound state of second order positon and one soliton with parameter values $\lambda_1 = 1+2.7 i $, $\lambda_3 =2+4.5i $ and $\alpha = 1 $. (a),(b) $s_{13} =s_{23} =s_{33} =1$; (c),(d) $s_{33} = 10$.}\label{5}
	\end{figure}

\begin{figure}[ht!]
		\centering
		\includegraphics[width=1\linewidth]{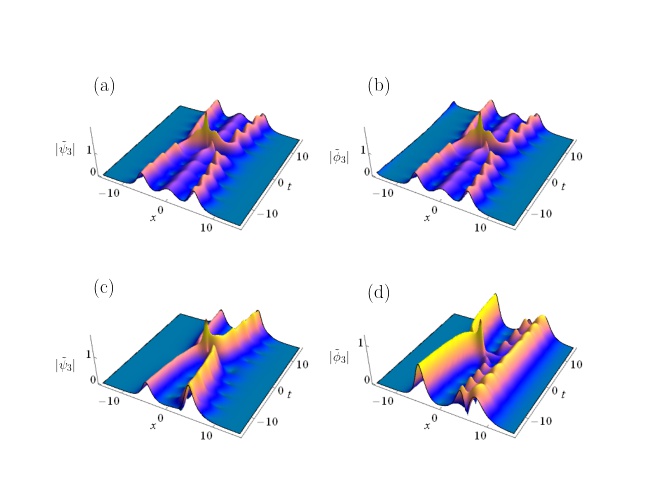}
		\caption{Bound state of second order positon and one soliton with parameter values same as Fig. 5 except for $\lambda_1 = 2+4.5 i $ and $\lambda_3 =1+2.7i $.}\label{6}
	\end{figure}

  In Figs. \ref{5} and \ref{6}, we bring out the interaction between bound states of  one soliton and second order positon. Figures \ref{5} (a) and \ref{5} (b) show the elastic interaction between bound states of one soliton with second order positon. If we increase the value of the parameter $s_{33}$, the bound state region moves in the forward direction towards $x$-axis with an increase in the amplitude of positon in $|\tilde\psi_3|$ and the amplitude of bound state region also increases in $|\tilde\phi_3|$ component, see Fig. \ref{5}(c) and \ref{5}(d). On the contrary, if we interchange the value of $\lambda_1$ and $\lambda_3$, the bound state region appears only on the second order positon which is evident from Fig. \ref{6} (a) and \ref{6} (b). We note here that in the previous case (Fig. 5) the bound states appear on the one soliton. When we increase the parameter value of $s_{33}$, the bound state region moves forward in $x$-axis along with an increase in amplitude of both soliton and positon, see Figs. \ref{6}(c) to \ref{6}(d). \\
	\\ \newline \textbf{Case 2: Interaction between second order positon and two solitons }\\
    \par
\begin{figure}[ht!]
		\centering
		\includegraphics[width=1\linewidth]{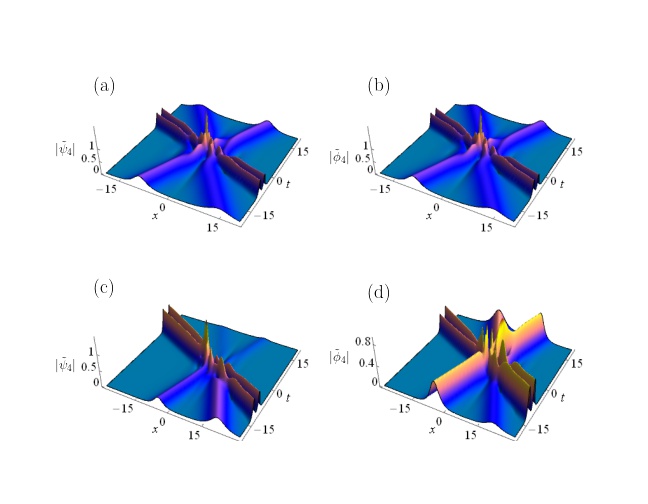}
		\caption{Interaction between second order positon and two soliton with parameter values $\lambda_1 = 13/10 +3/4~ i $, $\lambda_3 =-1/2+1/2 ~i $, $\lambda_4=1/2+1/3~ i $ and $\alpha = 2/9 $. (a),(b) $s_{14} =s_{24} =s_{34} =1$; (c),(d) $ s_{34} = 10$.}\label{7}
	\end{figure}

 To obtain second order positon - two soliton solution we take $N=4$ in Eq.\eqref{dt}.  Now the determinants of \eqref{dt} consists of four eigenvalues, $\lambda_k$, $k=1,2,3,4$.  By limiting the spectral parameter $\lambda_2 \rightarrow \lambda_1 + \epsilon$  with  $\lambda_3$ and $ \lambda_4$ remains unchanged, we can obtain a hybrid solution of second order positon and two soliton for the CH equation.  Since the explicit form of this solution is also very lengthy to present we illustrate the solution only pictorially.  The solution plots are given in Fig. \ref{7}. For the choice  $s_{14},s_{24},s_{34}=1$, one can see the occurrence of elastic interaction between two soliton and second order positon in Figs. \ref{7}(a) and \ref{7}(b). If we increase the parameter $s_{34}$, from Fig. \ref{7}(c), we can see that the amplitude of both  the two soliton and second order positon get suppressed in the $|\tilde\psi_4|$ component. On the other hand, the amplitude of $|\tilde\phi_4|$ component gets enhanced, see Fig.\ref{7}(d). Similarly, if we enhance the parameter value $s_{24}$, the amplitude gets enlarged in $|\tilde\psi_4|$ and suppressed in $|\tilde\phi_4|$ component.\\
	\newline
	\textbf{Case 3: Interaction between third order positon and one soliton}\\
	\par
\begin{figure}[ht!]
		\centering
		\includegraphics[width=1\linewidth]{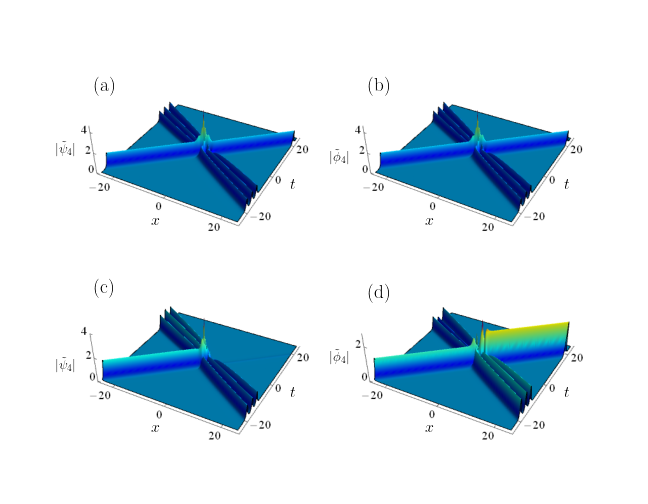}
		\caption{Interaction between third order positon and one soliton with parameter values $\lambda_1 = 13/20 +3/4 i $, $\lambda_4 =-3/4+i $, and $\alpha = 1/8 $. (a),(b) $s_{14} =s_{24} =s_{34} =1$; (c),(d) $s_{34} = 100$.}\label{8}
	\end{figure}

\begin{figure}[ht!]
		\centering
		\includegraphics[width=1\linewidth]{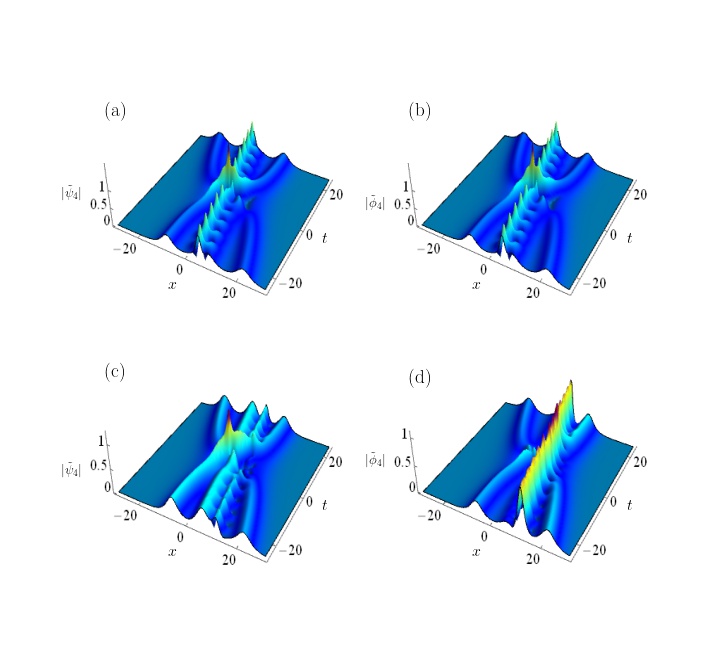}
		\caption{Bound state of third order positon and one soliton with parameter values $\lambda_1 = 1+ 1.9 i $, $\lambda_4 =1.9 + 3.9 i $, and $\alpha = 1$. (a),(b)$s_{14} =s_{24} =s_{34} =1$; (c),(d) $s_{34} = 100$.}\label{9}
	\end{figure}

\begin{figure}[ht!]
		\centering
		\includegraphics[width=1\linewidth]{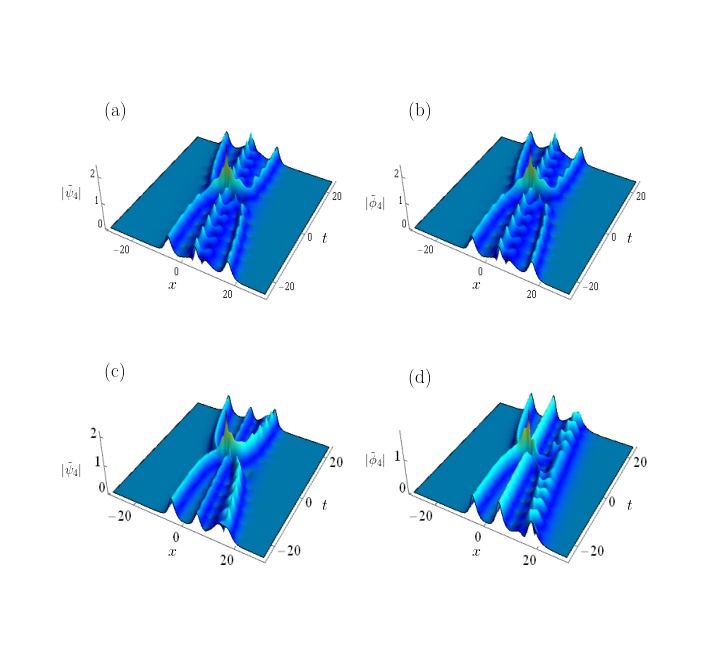}
		\caption{Bound state of third order positon and one soliton with parameter values  $\lambda_4 = 1+ 1.9 i $, $\lambda_1 =1.9 + 3.9 i $ and $\alpha = 1$. (a),(b) $s_{14} =s_{24} =s_{34} =1$; (c),(d) $ s_{34} = 10$.}\label{10}
	\end{figure}

 Next, we construct the hybrid solution of third order positon and one soliton solution.  To derive this solution we consider $N=4$ in Eq.\eqref{dt}.  To obtain third order positon we restrict the eigenvlues  $\lambda_{2,3} \rightarrow \lambda_1 + \epsilon$ and keep the eigenvalue $\lambda_4$ as it is. This choice provides the desired third order positon - one soliton solution. Figure \ref{8} represents the interaction between third order positon and one soliton on the zero intensity background. In Fig. \ref{8}(a) and \ref{8}(b), we can see that the elastic interaction between one soliton and third order positon exist for the choice $s_{14} =s_{24} =s_{34} =1$. From Fig. \ref{8}(c) and \ref{8}(d), we can confirm that inelastic interactions occurs as we increase the parameter values $s_{34}$. In Fig. \ref{8}(c), we can see that the one soliton disappears after the interaction and the third order positon remains unchanged in one component. However, the amplitude of third order positon and one soliton gets enhanced after the interaction with other component which is illustrated in Fig. \ref{8}(d). On the other hand, one positon get suppressed in $|\tilde{\phi}_4|$ component and enhanced in $|\tilde{\psi}_4|$ component while increasing the parameter value $s_{24}$. Figure \ref{9} shows the bound state among the one soliton and third order positon. If we increase the parameter values ($s_{24}$ or $s_{34}$), the bound state region moves forward in one component and the amplitude of the bound state region becomes higher in the other component. In Fig.\ref{10}(a) and \ref{10}(b), we can see that the bound states appear on the third order positon. As we have seen earlier, in cases 1 and 2, the bound state region moves forward while incrementing the parameter value $s_{24}$ or $s_{34}$. Here we come across an inelastic interaction between positon and soliton with a change in amplitude (see Fig. \ref{10}(c) and \ref{10}(d) ). 
	
	\begin{subequations}
 In the above subsections we derived positon-soliton type hybrid solutions of various forms. However, one can present a generalized form of the determinant formula for the hybrid  (positon - soliton) solution for the CH equations in the form
\begin{eqnarray}			\widehat{W}_{lN}&=&\left|\dfrac{\partial^{r_i}}{\partial\epsilon^{r_i}}\Big|_{\epsilon\rightarrow0}(W_{lN})_{ij}\right|_{3N\times3N},\\
r_i&=&\begin{cases}
\left[\frac{i-1}{3}\right], & i\leq 3m,	\\
0,& i>3m,
\end{cases}
\end{eqnarray}
\end{subequations}
where $l=1,2,3$ and $[i]$ is the floor function and $N= m + n$. Here, $m$ and $n$ corresponds to the evolution of positon and soliton, respectively. By choosing different values for $m$ and $n$ we can construct hybrid solutions composed of solitons and positons. For example, the choice $m=2$ and $n=1$ will give rise to second order positon and one soliton solution and if we take $m=2$ and $n=2$, we can get second order positon and two soliton solution. For the choice $n=0$ ($N=m$), the hybrid solution transforms into positon solution of CH equation. Earlier, to derive positon solutions, we expanded all the eigenfunctions in Taylor series with respect to $\epsilon$ but in the hybrid solution we differentiate only the eigenfunctions that correspond to the value of $m$ while leaving the other eigenfunctions unchanged. In this way, one can obtain the hybrid solutions for different values of $m$ and $n$ which satisfies the Eq. \eqref{1}. 
	\section{Solutions of CH equation on plane wave background: Breather Positons}
    It is well known that the breathers of NLS type equations can be constructed on plane wave background in DT method.  Breather positons can also be derived using plane wave background using GDT method.  In this section, we construct this solution and study the interaction dynamics between positons and breather positons.  We take seed solution in the form
	\begin{equation}
		\psi(x,t) = c_1 e^{k_1 x+ \omega_1 t}, \qquad \phi(x,t) = c_2 e^{k_2 x+ \omega_2 t},\label{ss}
	\end{equation}
	where $c_1$, $c_2$, $k_1$, $k_2$, $\omega_1$ and $\omega_2$  are real parameters.  Substituting the above expressions in Eq.\eqref{e1}, we can obtain the following two dispersion relations, namely
	\begin{subequations}
		\begin{eqnarray}
			\omega_1 = \dfrac{-k_1^2}{2}+ c_1^2 +c_2^2 + \alpha\left[k_1^3 - k_1(6c_1^2 + 3c_2^2)-3c_2^2k_2\right],\\
			\omega_2 = \dfrac{-k_2^2}{2}+ c_1^2 +c_2^2 + \alpha\left[k_2^3 - k_2(6c_2^2 + 3c_1^2)-3c_1^2k_1\right].	
		\end{eqnarray}
	\end{subequations}
	\par 	Solving the Lax pair equation \eqref{lp} with the seed solutions \eqref{ss} is quite cumbersome. So we adapt the following procedure to solve the Lax pair equation \eqref{lp}. The spatial part of the Lax pair equation associated with the above seed solutions \eqref{ss} can be represented as
	\begin{subequations}\label{ode}
		\begin{eqnarray}
			u_{1x}&=& -\dfrac{i\lambda}{6 \alpha} u_1 - c_1 e^{i(k_1 x+ \omega_1 t)} v_1 - c_2 e^{i(k_2 x+ \omega_2 t)}w_1,\\
			v_{1x}&=& c_1 e^{-i(k_1 x+ \omega_1 t)} u_1 +\dfrac{i\lambda}{12 \alpha} v_1,\\
			w_{1x}&=& c_2 e^{-i(k_2 x+ \omega_2 t)} u_1 +\dfrac{i\lambda}{12 \alpha} w_1,		  
		\end{eqnarray}
	\end{subequations}
	\par To solve the system of first order ODEs \eqref{ode}, we recaste them in a single third order linear differential equation in the variable $u_1(x,t)$ and its derivatives. The resultant ODE takes the following form
 \begin{eqnarray}
		&&u_{1xxx} +u_{1x}\left[c_1^2 +c_2^2 +\dfrac{\lambda}{6 \alpha}\left(k_2 + \dfrac{\lambda}{12 \alpha}\right)+ \left(k_1 + \dfrac{\lambda}{12 \alpha}\right)\left(-k_2 + \dfrac{\lambda}{12 \alpha}\right)\right] - i u_{1xx} \notag \\
		&&\times(k_2+k_1)- i u_{1}\left[c_1^2 (k_2-k_1) + \left(\dfrac{\lambda}{12 \alpha}+ k_1\right)\left(c_1^2 +c_2^2 + \dfrac{\lambda}{6 \alpha}\left(k_2 + \dfrac{\lambda}{12 \alpha}\right)\right)\right]=0.\qquad 
	\end{eqnarray}

 By integrating this third order differential equation, we can get the explicit form of $u_1(x,t)$ and from that we can derive the other two functions $v_1$ and $w_1$. The general solution of the third orde ODE relies on the roots of the characteristics equation, 
	\begin{eqnarray}
		a^3 - i a^2(k_2+k_1)+a\left[c_1^2 +c_2^2 +\dfrac{\lambda}{6 \alpha}\left(k_2 + \dfrac{\lambda}{12 \alpha}\right)+ \left(k_1 + \dfrac{\lambda}{12 \alpha}\right)\left(-k_2 + \dfrac{\lambda}{12 \alpha}\right)\right] \notag \\
		- i\left[c_1^2 (k_2-k_1) + \left(\dfrac{\lambda}{12 \alpha}+ k_1\right)\left(c_1^2 +c_2^2 + \dfrac{\lambda}{6 \alpha}\left(k_2 + \dfrac{\lambda}{12 \alpha}\right)\right)\right]=0,\qquad \label{cp} 
	\end{eqnarray} 
	which is the eigenvalue equation of $u_1(x,t)$. The cubic polynomial \eqref{cp} given above admits one of the three  types of roots, namely (i) three equal roots, (ii)  two equal and one different roots and (iii) three different roots. The nonlinear localized solutions, namely breathers, rogue waves and dark solitons can be derived by appropriately choosing one of these cases \cite{vp2}. As far as the cases (i) and (ii) are concerned, we obtain a condition on $\lambda$ (critical eigenvalue) and it leads to rogue wave solution. Since positon solution shares same eigenvalue, we can identify them only through the case (iii).  As we are interested in studying the interaction dynamics between positon and breather positon solutions we consider only the case (iii).  Hence the general solution of Eq.\eqref{cp} with three different roots can be expressed in the following form
	\begin{equation}
		u_1= R_1(t)e^{a_1 x} + R_2 (t) e^{a_2 x} + R_3(t) e^{a_3 x}, \label{rt}
	\end{equation}
	where $a_1, a_2$ and $a_3$ are three different roots of the cubic polynomial and $R_1(t), R_2(t)$ and $R_3(t)$ are arbitrary functions of $t$ whose explicit forms can be obtained by substituting  Eq.\eqref{rt} into the temporal part of the Lax pair equations and integrating the resultant equations. Doing so, we find
	\begin{subequations}
		\begin{eqnarray}
			R_j = exp\left[ \left(A_1 + \dfrac{B_1 L_j}{i (k_2 - k_1)} + B_2 \left(-a_j - \dfrac{i \lambda}{6 \alpha}+ \dfrac{iL_j}{(k_2 -k_1)}\right)\right)t\right], j=1,2,3.\notag \\
		\end{eqnarray}
		with 
		\begin{eqnarray}
			A_1& = & - \dfrac{i\lambda^2}{48 \alpha^2} - \dfrac{i\lambda^3}{96 \alpha^2} + \dfrac{i}{2}(c_1^2 + c_2^2)(1+ \lambda/2) - 2i\alpha(c_2^2 k_2 + c_1^2 k_1 ),\notag\\
			B_1 & = & - \dfrac{\lambda^2}{16 \alpha} - \dfrac{\lambda}{8 \alpha} + \dfrac{k_1 \lambda}{4} + \dfrac{k_1}{2} + \alpha ( 2(c_1^2 + c_2^2)- k_1^2),\notag \\
			B_2&=& - \dfrac{\lambda^2}{16 \alpha} - \dfrac{\lambda}{8 \alpha} + \dfrac{k_2 \lambda}{4} + \dfrac{k_2}{2} + \alpha ( 2(c_1^2 + c_2^2)- k_2^2),\notag \\
			L_j & = & a_j^2 + \left(\dfrac{i \lambda}{12 \alpha}- i k_2 \right)a_i + c_1 ^2 +c_2^2 + \dfrac{\lambda}{6 \alpha}\left(\dfrac{\lambda}{12 \alpha}+ k_2\right),
		\end{eqnarray}
		
		where
		\begin{eqnarray}
			a_1 & =&	\frac{1}{3} i ({k_1}+{k_2})-\bigg(2^{1/3} \bigg(3 {c_1}^2+3 {c_2}^2+{k_1}^2-{k_1} {k_2}+{k_2}^2+\frac{{k_1}
				{\lambda_1}}{4 \alpha }+\frac{{k_2} {\lambda_1}}{4 \alpha }\notag\\&&+\frac{{\lambda_1}^2}{16 \alpha ^2}\bigg)\bigg)\bigg/
			\bigg(3 \bigg(A_{11}+
			\sqrt{ (4 (A_{12})^3+
				(A_{11})^2)}\bigg)^{1/3}+
			\frac{1}{3\times 2^{1/3}}
			\bigg(A_{11}\notag\\&&+
			\sqrt{ (4 (A_{12})^3+
				(A_{11})^2)}\bigg)^{1/3}\bigg),
		\end{eqnarray}
		\begin{eqnarray}
			a_2 &=&\frac{1}{3} i ({k_1}+{k_2})+\bigg((1+i \sqrt{3}) A_{12}\bigg)\bigg/
			\bigg(3\times 2^{2/3} 
			\bigg(A_{11}+
			\sqrt{ (4 (A_{12})^3+
				(A_{11})^2)}\bigg)^{1/3}\notag\\&&-
			\frac{1}{6 \times 2^{1/3}}(1-i \sqrt{3}) 
			\bigg(A_{11}+
			\sqrt{ (4 (A_{12})^3+
				(A_{11})^2)}\bigg)^{1/3}\bigg),
		\end{eqnarray}
		\begin{eqnarray}
			a_3 & = & \frac{1}{3} i ({k_1}+{k_2})+\bigg((1-i \sqrt{3}) A_{11}\bigg)\bigg/
			\bigg(3\times 2^{2/3} 
			\bigg(A_{11}+
			\sqrt{ (4 (A_{12})^3+
				(A_{11})^2)}\bigg)^{1/3}\notag\\&&-
			\frac{1}{6\times 2^{1/3}}(1+i \sqrt{3}) 
			\bigg(A_{11}+
			\sqrt{ (4 (A_{12})^3+
				(A_{11})^2)}\bigg)^{1/3}\bigg).	
		\end{eqnarray}
		with
		\begin{eqnarray}
			A_{11}&=&-9 i {c_1}^2 {k_1}+18 i {c_2}^2 {k_1}-2 i {k_1}^3+18 i {c_1}^2 {k_2}-9 i {c_2}^2 {k_2}+3 i
			{k_1}^2 {k_2}+3 i {k_1} {k_2}^2 \notag\\&&-2 i {k_2}^3+\frac{9 i {c_1}^2 {\lambda}}{4 \alpha }+\frac{9 i {c_2}^2 {\lambda}}{4 \alpha }-\frac{3 i {k_1}^2 {\lambda}}{4 \alpha }+\frac{3 i {k_1} {k_2} {\lambda}}{\alpha }-\frac{3 i {k_2}^2
				{\lambda}}{4 \alpha }+	\frac{3 i {k_1} {\lambda}^2}{16 \alpha ^2}\notag\\&&
			+\frac{3 i {k_2} {\lambda}^2}{16 \alpha ^2}+\frac{i {\lambda}^3}{32
				\alpha ^3},\\	
			A_{12} &=& 3 {c_1}^2+3 {c_2}^2+{k_1}^2-{k_1} {k_2}+{k_2}^2+\frac{{k_1} {\lambda}}{4 \alpha }+\frac{{k_2}
				{\lambda}}{4 \alpha }+\frac{{\lambda}^2}{16 \alpha ^2}.
		\end{eqnarray}
		
	\end{subequations}
	Substituting the above functions back into the general solution \eqref{rt}, we can get the explicit expression for the function $u_1$. From that, we will obtain the other two eigenfunctions $v_1$ and $w_1$ in the form,
	\begin{eqnarray}
		v_1 &=& \dfrac{ e^{-i(k_1 x+ \omega_1 t)}}{i c_1(k_2 - k_1)} \left[u_{1xx} + i\left(\dfrac{\lambda}{12 \alpha} - k_2\right)u_{1x}+ u_1\left(c_1^2 + c_2^2 +\dfrac{\lambda}{6 \alpha}\left(\dfrac{\lambda}{12 \alpha}+ k_2\right)\right)\right], \notag \\
		w_1 &=& \dfrac{ e^{-i(k_2 x+ \omega_2 t)}}{k_2}\left[- u_{1x}- \dfrac{i \lambda}{6 \alpha} u_1 - c_1 e^{i(k_1 x+ \omega_1 t)} v_1\right]. \label{ef2}
	\end{eqnarray}
	With the help of these eigenfunctions \eqref{rt} and \eqref{ef2}, we can construct the breather solution by substituting them in the DT formulas \eqref{dt}. In the following subsection, we will discuss the method of constructing the degenerate breathers or breather positon from the obtained eigenfunctions in nontrivial background and investigate the interaction between them.

	\subsection{Interaction dynamics of breather positons and positons}
	To construct the degenerate nonlinear wave solution, we substitute the eigenfunctions \eqref{rt} and \eqref{ef2} in the GDT formula \eqref{dt1} and expand the eigenfunction in Taylor series at $\epsilon$ by limiting the spectral parameter as $\lambda_2 = \lambda_1 + \epsilon$. The explicit expression of the resultant solution is quite lengthy. Hence, we present only the plots of the functions $|\psi_2|$ and $|\phi_2|$. Here, we investigate two cases. In the first case, we consider both the components having plane wave seed solutions and in the other case we consider only one of the components having plane wave seed solution.\\

	\par
\begin{figure}[ht!]
		\centering
		\includegraphics[width=1\linewidth]{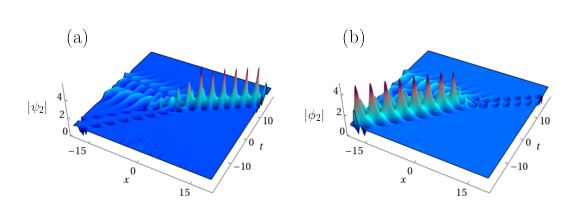}
		\caption{Inelastic collision of second order b-p with (a) four-petaled second order b-p and evolves as second order b-p in $|\psi_2|$ component and (b) second order b-p which transits into four-petaled second order b-p in $|\phi_2|$ component for the parameter values $\lambda_1 = 1.2 i $, $c_1 =c_2 =1, k_1 = -0.9, k_2 = 0.9$ and $\alpha= 0.1$.}
		\label{12}
	\end{figure}

 \textbf{Collision between two b-ps:} We consider both the components $\psi_2$ and $\phi_2$ having plane wave seed solution $c_1\neq0$ and $c_2 \neq 0$ in \eqref{ss} which induce the evolution of breathers. In this case, we can see that the second order b-p and second order four-petaled  b-p collide and they become a second order b-p with an increase in amplitude in the $|\psi_2|$ component, see Fig.~\ref{12}(a). In the $|\phi_2|$ component, one can observe that the second order b-p with different amplitude collide and they become four-petaled second order b-p, see Fig.~\ref{12}(b).  The occurrence of this inelastic collision between different nonlinear waves may find applications in nonlinear optics particularly in designing optical switches.\\
	
	\par
\begin{figure}[ht!]
		\centering
		\includegraphics[width=1\linewidth]{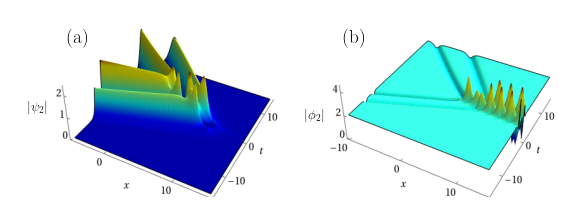}
		\caption{(a) Evolution of second order degenerate bright soliton in $|\psi_2|$ component is reflected by b-p pulses, (b) Second order degenerate dark soliton in $|\phi_2|$ component is reflected by the second order b-p for the parameter values $\lambda_1 = i$, $c_1 =0.001, c_2 =2, k_1 = -0.2, k_2 = 0.2$ and $\alpha= 0.1$.}
		\label{11}
	\end{figure}

 \textbf{Collision between b-p and positon:} Next, we choose one of the components as a plane wave solution and the other component close to zero. The choice $c_1\rightarrow0$  enforces the raise of bright and dark soliton in  $|\psi_2|$ and  $|\phi_2|$ component, respectively and the choice $c_2 \neq 0$ creates the evolution of breathers in both the components. From Fig.~\ref{11}(a), we can observe that the second order degenerate bright soliton is reflected back by b-p pulses in the $|\psi_2|$ component whereas in the $|\phi_2|$ component second order degenerate dark soliton is reflected by the second order b-p  which is demonstrated in Fig.~\ref{11}(b).

	
	\section{Conclusion}
	In this paper, to begin with,  we have constructed higher order positon solutions for the Eq.\eqref{e1} through generalized Darboux transformation method with zero seed solution. Upon enhancing the value of higher order nonlinear coefficient ($\alpha$), we have demonstrated that not only the width of positon increases but also the distance between the two waves increases. We observe that this property is not only in the second order positon solution but also present in the higher order positons, say third and fourth order positon solutions. Further, we have studied the asymptotic nature of the second order positon solution of Eq. \eqref{e1}. By comparing the absolute maximal amplitude of one soliton solution with the shifted second order positon, we derived an expression for the time-dependent phase shift which is directly proportional to the system parameter ($\alpha$). With the time dependent displacement, the second order positon solution can be viewed as  two `one-solitons' in their asymptotic limits since it travels as a single component at smaller times. In positons, no energy exchange happens during collision. We have also constructed hybrid solutions composed of solitons and positons for the Eq.\eqref{e1} using generalized Darboux transformation method. Elastic and inelastic collision of higher order soliton with positon along with its bound states are investigated in detail and the outcomes are demonstrated graphically. Here, the positon while colliding with a soliton exchanges  energy at certain parametric values. We have also constructed the degenerate second order breather-positon solutions with nontrivial background for the Eq.\eqref{e1}. Further, we have studied the dynamics of the underlying solutions by considering two cases. In the first case, we have considered both the components  to have plane wave solution as background. In this case we have shown that the inelastic collision exists between second order breather-positon and  four-petaled second order breather-positon in the components $|\psi_2|$ and  $|\phi_2|$.  In the second case, we have considered one component to have plane wave background and other component to have zero background. Here, we have shown that the evolution of second order degenerate bright soliton is  reflected back by the breather-positon pulses in the component $|\psi_2|$ and the second order dark soliton is reflected back by the second order breather-positon in the $|\phi_2|$ component. 
 \par As we pointed out earlier, deriving breather positon solutions involve tedious calculations. Hence in this manuscript, we have restricted our efforts only on deriving  first order breather positon solution.  In future, we intend to construct higher order breather-positon solutions of coupled Hirota equations.  We will also derive hybrid solutions of breather-positon solutions and study the interaction dynamics of breathers and breather-positons in near future. Further, we plan to analyze these solutions using deep learning methods in an upcoming work.  The study on the interaction of different nonlinear waves has numerous applications in nonlinear optics and our results will be helpful in such problems. \\ 
	
	\section*{Data Availability}
	The data that support the findings of this study are available within this article.
	\section*{Competing Interests}
	The authors have no competing interests.
	\section*{Funding Statements}
	SM wishes to thank MoE-RUSA 2.0 Physical Sciences, Government of India for providing a fellowship to carry out this work. NVP thanks the Department of Science and Technology (DST), India for the financial support under Women Scientist Scheme-A. The work of MS was supported by DST-SERB, Government of India, under the Grant No. CRG/2021/002428.

  \section*{Authors Contributions Statement}
	 All the authors have contributed equally to the research and to the writing up of the paper.

  \bibliographystyle{elsarticle-num}

\end{document}